\documentclass{ws-procs9x6}

\newcommand{\refeq}[1]{(\ref{#1})}
\def\etal {{\it et al.}}
\def\eg{{e.g., }}

\begin{document}

\title{REDSHIFT ANOMALIES WITH UNIVERSAL FREE-FALL}

\author{M.A.\ HOHENSEE$^*$ and H.\ M\"ULLER}

\address{Department of Physics, University of California\\
Berkeley, CA 94720, USA\\
$^*$E-mail: hohensee@berkeley.edu}

%

\begin{abstract}
For most theories which parametrize modifications of General Relativity, including those which violate the equivalence principle, gravitational redshift tests typically offer weaker constraints on such test parameters than do precision measurements of the universality of free fall (UFF) and local Lorentz invariance (LLI).  Although redshift anomalies are often linked with violations of UFF or LLI, they do not have to be.  We offer a simple model in which particle masses anomalously vary with the gravitational potential.  This generates gravitational redshift anomalies unconstrained by existing tests of UFF or LLI.  We propose new experiments to limit such effects.
\end{abstract}

\bodymatter

\section*{}

\vskip -10pt


The gravitational redshift is a classic test of General Relativity (GR).  It was the first test that Einstein proposed for GR~\cite{Einstein:1911}, and its verification by Pound, Rebka, and Snider in the early 1960s~\cite{Pound:1960} was the first of many increasingly precise tests of Einstein's equivalence principle (EEP) and GR~\cite{Will:2006}.  The precision of some tests, particularly those of the universality of free fall (UFF)~\cite{Schlamminger:2008}, is such that their indirect limits~\cite{Will:2006,MuellerICAP} on redshift anomalies are more precise than current tests~\cite{Vessot:1980,Cacciapuoti:2009}.  Modern redshift tests can nevertheless bound anomalies that are presently poorly constrained.  We present a simple test Lagrangian which can describe such anomalies, and show that placing an atom interferometer (AI) on a sounding rocket could improve terrestrial bounds on such anomalies by a factor of more than $10^{10}$, and by at least a factor of 10 over solar system tests of a particular generalization.  
The EEP is a cornerstone of GR, and has been subjected to variety of experimental~\cite{Pound:1960,Schlamminger:2008,datatables} and astrophysical tests~\cite{Will:2006}.   EEP requires that: i) a body's gravitational mass always equals its inertial mass --- known as UFF, or the weak equivalence principle; ii) local Lorentz invariance (LLI): the results of nongravitational experiments do not depend upon the velocity of the freely falling frame in which they are performed; and iii) local position invariance (LPI): the result of a nongravitational experiment does not depend upon where and when it is performed~\cite{Will:2006}.  These have been experimentally tested by i) torsion balance experiments~\cite{Schlamminger:2008}, ii) a wide variety of Lorentz symmetry tests~\cite{datatables}, and iii) gravitational redshift experiments~\cite{Pound:1960,HafeleKeating,Vessot:1980,Mueller:2010}.  The three components of the EEP are closely interrelated, and follow from UFF for most theories of gravity~\cite{Schiff:1960,Will:2006}.  

\begin{figure}[t]
\begin{center}
\psfig{file=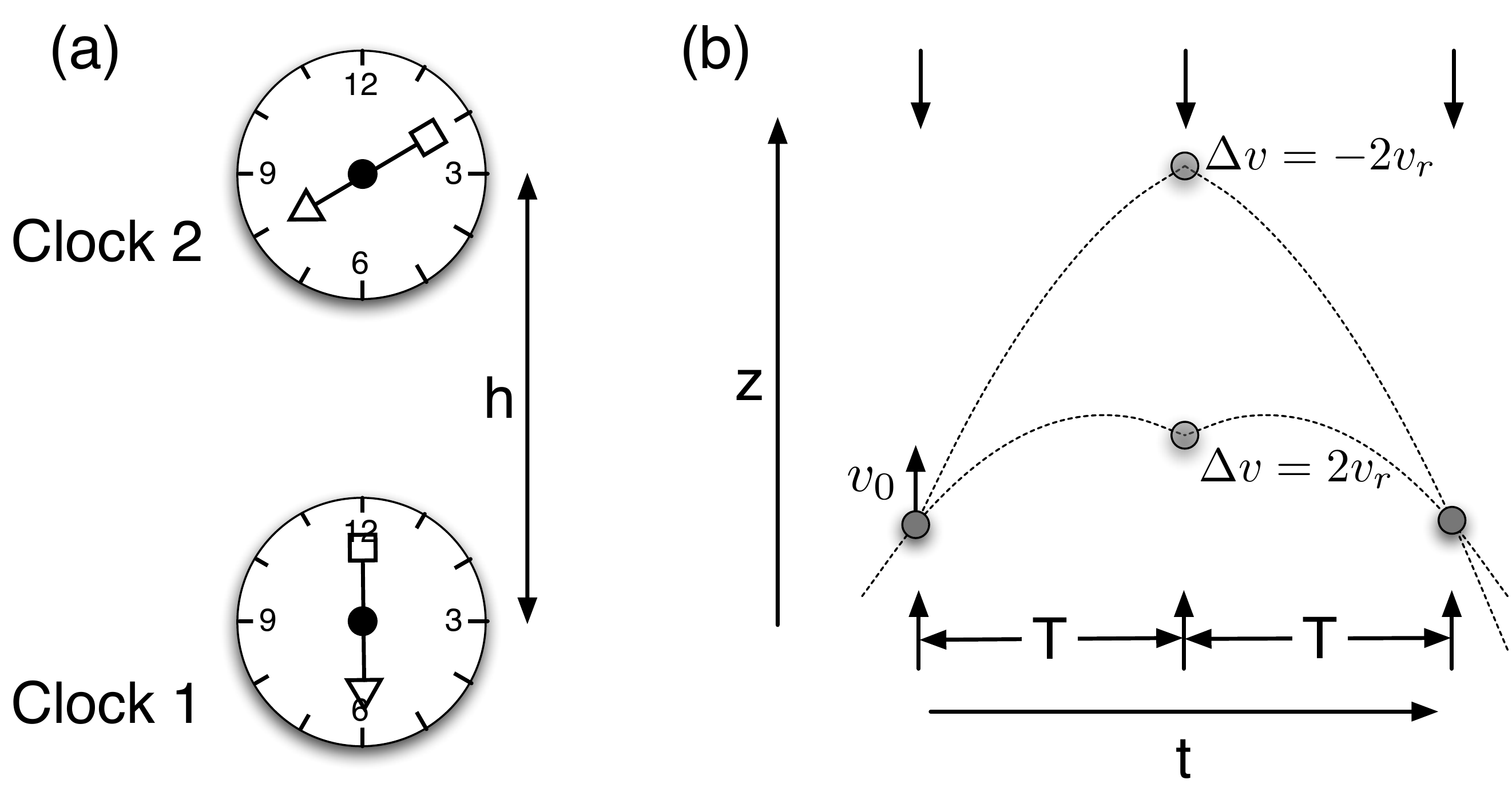,width=3in}
\end{center}
\caption{Two redshift experiments. (a) An unmoving separated clock experiment similar to the Pound-Rebka-Snider experiment.  Two clocks are separated by a distance $h$ along the gravitational potential gradient.  Leading order redshift anomalies appear in the ratio $\nu_{2,1}/\nu_{0}$ as $1+\frac{gh}{c^{2}}(1+\beta)$.  (b) An AI redshift test.  Leading order anomalies appear as an interferometer phase $\Delta\phi=(1+\beta)kgT^{2}$.}
\label{fig:clocksep}
\end{figure}

A simple redshift experiment involves two clocks $1$ and $2$ which locally tick at frequency $\nu_{0}$, separated by a distance $h$ as in Fig.\ \ref{fig:clocksep}.  If we neglect the Earth's rotation and use the Schwarzschild solution to the Einstein field equations, the leading order gravitational redshift is
\begin{equation}
\nu_{2,1}=\nu_{0}\left(1+\frac{gh\left(1+\beta_{\rm I}\right)}{c^{2}}\right),\label{eq:redI}
\end{equation}
where $\nu_{2,1}$ is the frequency of clock $2$ as measured by clock $1$, and $\beta_{\rm I}$ is the leading order redshift anomaly, if any. Equation \refeq{eq:redI} is obtained from the effective nonrelativistic point-particle Lagrangian
\begin{equation}
L=mc^{2}\left(1+\frac{gz}{c^{2}}\left(1+\beta_{\rm I}\right)-\frac{\dot{z}^{2}}{2c^{2}}\right).\label{eq:lagI}
\end{equation}
If $\beta_{\rm I}$ is a universal constant, this Lagrangian produces an unobservable rescaling of $g$, since any local measurement of the acceleration of free fall would yield $g'=g(1+\beta_{\rm I})$.  The anomaly is observable if it varies for different systems, \eg if $(1+\beta_{\rm I})$ in the clocks' Lagrangian is replaced by $(1+\beta_{\rm I}+\beta_{\rm II})$.  
In terms of $g'$, the observed anomaly would be 
\begin{equation}
\nu_{2,1}=\nu_{0}\left(1+\frac{g'h\left(1+\beta_{\rm II}\right)}{c^{2}}\right),
\end{equation}
and may therefore be constrained by gravitational redshift tests.  This effect can also be constrained by torsion balance tests of the UFF, since it predicts a different acceleration of free fall for different systems.  The best limits on UFF violations are presently obtained by the E\"ot-Wash experiment~\cite{Schlamminger:2008}, which bounds the E\"otv\"os parameter $|\eta_{\rm Be,Ti}|\leq1.8\times10^{-13}$ for beryllium relative to titanium.  Using Haugan's~\cite{Haugan:1979} formula to estimate the resultant limits for magnetic hyperfine transition energies, this bounds the observable atomic clock $\beta_{\rm II}$ at the level of $2\times 10^{-7}$, more than two orders of magnitude smaller than the accuracy of current tests~\cite{Vessot:1980}.

Redshift anomalies can also be measured by AIs, which use matter waves to measure the proper time difference for atoms moving along different paths in a gravitational potential~\cite{Borde:2008}.  In its simplest form, the AI is essentially a Hafele-Keating~\cite{HafeleKeating} separated clock comparison test, in which the clocks follow two different ballistic trajectories.  As outlined in more detail elsewhere~\cite{Mueller:2010,MuellerTheseProceedings}, atoms are launched vertically with velocity $v_{0}$ at time $t=0$, and undergo Bragg scattering from the standing wave generated by a retroreflected laser, synchronizing their `clocks.'   The state of each atom has equal amplitude to have $v_{z}=v_{0}$ or $v_{z}=v_{0}+2v_{r}$, where $v_{r}=\hbar k/m$ is the atoms' recoil velocity; thus the atoms are coherently split to travel different paths.  After a time $T$, atoms on the two paths are given equal and opposite momentum kicks of $2\hbar k$, so that they overlap and can be interfered with one another at time $2T$ with a second Bragg pulse.  We find the relative phase of the matter waves along the two arms of the interferometer by measuring the number of atoms in each of the two possible final momentum states.  Since matter waves oscillate at the Compton frequency ($mc^{2}/(2\pi\hbar)\sim 3.2\times10^{25}$ Hz, for Cs atoms), very small changes in the proper-time along one arm relative to the other can be resolved.  For the test models described above, the total interferometer phase is the sum of the free evolution phase $\Delta\phi_{\rm free}$ and the interaction phase $\Delta\phi_{\rm int}$: $\Delta\phi=(1+\beta)kgT^{2}$~\cite{Borde:2008,Mueller:2010,MuellerTheseProceedings}.  As before, if the local $g$ is measured independently of the AI, only differences in the value of $\beta$ that also give rise to violations of UFF can generate redshift anomalies.

More subtle effects can appear at higher order.  Consider a model where particles' masses anomalously vary in their local gravitational potential
\begin{equation}
L=mc^{2}+mc^{2}\left(1+\beta_{\rm III}\frac{gz}{c^{2}}\right)\left(\frac{gz}{c^{2}}-\frac{\dot{z}^{2}}{2c^{2}}\right)\label{eq:lagIII}.
\end{equation}
This model includes terms resembling those appearing at higher order in the expansion of the Schwarzschild metric.\footnote{$\beta_{\rm III}\neq 0$ may be understood as several anomalous terms in the expansion.  Higher order gradients and GR must be considered when experiments are sensitive to $\beta\sim 6$.}  Whether the resulting redshift anomalies are violations of the EEP, or merely corrections to GR is beyond our present scope.  This Lagrangian modifies point-particle kinematics: the local acceleration of free fall is, to first order in $\beta_{\rm III}$,
\begin{equation}
\ddot{z}=-g\left(1+\beta\left[\frac{gz}{c^{2}}+\frac{\dot{z}^{2}}{2c^{2}}\right]\right),
\end{equation}
and is the same for all objects.  Expanding to $O(T^{2})$, and assuming that photon momenta scale like the masses with height,\footnote{Dropping this assumption changes the second term by a factor of $3$.} the total  AI phase is
\begin{equation}
\Delta\phi = kgT^{2}-\beta_{\rm III}\frac{1}{2}\frac{v_{0}^{2}}{c^{2}}kgT^{2}-\beta_{\rm III}\frac{v_{0}v_{r}}{c^{2}}kgT^{2}.\label{eq:typeIIIphase}
\end{equation}
The second term in \refeq{eq:typeIIIphase} is similar to one derived by Dimopoulos \etal~\cite{Dimopoulos:2008a} using GR for a differently configured AI.  The third term is absent from that analysis.
The most sensitive tests of the gravitational redshift in AIs are precise to 7 ppb~\cite{Mueller:2010}, which in light of the small launch velocity ($v_{0}=1.53$ m/s), translates to a bound of $|\beta_{\rm III}|\leq 5\times 10^{8}$.  
This sensitivity is not unique to the AI.  Moving Pound-Rebka-Snider tests with $\vec{v}\cdot{\vec{g}}/|\vec{g}|=v_{0}\ll c$ observe
\begin{equation}
\nu_{2,1}=\nu_{0}\left(1+\frac{gh}{c^{2}}\left[1-\beta_{\rm III}\frac{v_{0}}{c}\left(1-\frac{v_{0}}{c}\right)\right]\right),
\end{equation}
where we ignore higher order terms in the Schwarzschild expansion which become relevant when $\beta_{\rm III}$ becomes as small as $1$. Motion of the Earth in the Sun's gravitational potential would permit Earth-stationary Pound-Rebka-Snider experiments with $1\%$ accuracy to bound $|\beta_{\rm III}|<10^{8}$ by repeated measurements made while the Earth moves between orbital perihelions.  

The effects of $\beta_{\rm III}$ are suppressed by one or more factors of $v_{0}/c$, the characteristic velocity of the apparatus along the gravitational potential gradient.  Limits on $\beta_{\rm III}$ can improved by carrying out an AI redshift test on a sounding rocket~\cite{Reasenberg:2010}.  The precision of the AI redshift test improves during the rocket's time in free fall, as the leading order AI signals vanish in free fall, while anomalies do not.  We estimate that $10^{-14}$ precision could be achieved by such an experiment~\cite{MuellerTheseProceedings}. At the same time, $(v_{0}/c)^{2}$ increases by a factor of up to $1.5\times10^{7}$ relative to the stationary test~\cite{Vessot:1980}.  Such a test could have a sensitivity to $\beta_{\rm III}\sim 1\times 10^{-4}$.  Note that such a test would also be sensitive to higher order GR effects not accounted for here~\cite{Dimopoulos:2008a}.

A generalization of this model is constrained by solar system tests of GR.  One possible generalization of Eq.\ \refeq{eq:lagIII} to a 2-body problem might be
\begin{equation}
L=\left[\frac{1}{2}\left(M\dot{\vec{R}}^{\,2}+\mu\dot{\vec{r}}^{\,2}\right)+\frac{GM\mu}{r}\right]\left(1+\beta_{\rm III}\frac{GM}{rc^{2}}\right),\label{eq:3d}
\end{equation}
where $M=m_{1}+m_{2}$ is the total mass, $\mu=m_{1}m_{2}/M$ is the reduced mass, and $\vec{R}$ and $\vec{r}$ are the center of mass and relative position vectors.
In the limit $\mu\ll M$, $z=r-r_{\oplus}\ll r_{\oplus}$, with $r_{\oplus}$ the Earth's radius, we recover the one-body Lagrangian \refeq{eq:lagIII}. Equation \refeq{eq:3d} causes anomalous periapsis precession, $\bar{\dot{\omega}}=\beta_{\rm III}\bar{\dot{\omega}}_{\rm GR}$, where $\bar{\dot{\omega}}_{\rm GR}$ is the GR precession.  The perihelion precession of Mercury fractionally differs from that predicted by GR by less than $1\times 10^{-3}$~\cite{Will:2006}, implying $|\beta_{\rm III}|\leq 1\times10^{-3}$.  This constraint may not apply to terrestrial AI tests, as $\beta_{\rm III}$ might vanish at large $r$ in another model.

\vskip -10pt

\section*{Acknowledgments}
We thank Juna Kollmeier, Jim Phillips, Nan Yu, Alan Kosteleck\'y and Jay Tasson for useful discussions.  This work is supported by the David and Lucile Packard Foundation, the Alfred P. Sloan Foundation, and a Precision Measurement Grant No. 60NANB9D9169 from NIST.


\begin{thebibliography}{xx}
\bibitem{Einstein:1911}
A.\ Einstein,
Annalen der Physik {\bf 35}, 898 (1911).

\bibitem{Pound:1960}
R.V.\ Pound and G.A.\ Rebka\ Jr.,
Phys. Rev. Lett. {\bf 4}, 337 (1960);
R.V.\ Pound and J.L.\ Snider,
Phys. Rev. Lett. {\bf 13}, 539 (1964);
Phys. Rev. {\bf 140}, B788 (1965).

\bibitem{Will:2006}
C.M.\ Will,
Living Rev. Rel. {\bf 9}, 3 (2006).

\bibitem{Schlamminger:2008}
S.\ Schlamminger \etal,
Phys.\ Rev.\ Lett. {\bf 100}, 041101 (2008).

\bibitem{MuellerICAP}
H.\ M\"uller,
to appear in Proc. 22nd Int. Conf. At. Phys. (2010).

\bibitem{Cacciapuoti:2009}
L.\ Cacciapuoti and Ch.\ Salomon,
Eur.\ Phys.\ J.\ Spec.\ Top. {\bf 172}, 57 (2009).

\bibitem{Vessot:1980}
R.F.C.\ Vessot \etal,
Phys.\ Rev.\ Lett. {\bf 45}, 2081 (1980);
R.F.C.\ Vessot,
Proc. IEEE {\bf 79}, 1040 (1991).

\bibitem{datatables}
V.A.\ Kosteleck\'y and N.\ Russell,
arXiv:0801.0287v3, (2010).

\bibitem{HafeleKeating}
J.C.\ Hafele and R.E.\ Keating,
Science {\bf 177}, 168 (1972).

\bibitem{Mueller:2010}
H.\ M\"uller, A.\ Peters, and S.\ Chu,
Nature {\bf 463}, 926 (2010).

\bibitem{Schiff:1960}
L.I.\ Schiff,
Am. J. Phys. {\bf 28}, 340 (1960).

\bibitem{Haugan:1979}
M.P.\ Haugan,
Ann. Phys. {\bf 118}, 156 (1979).

\bibitem{Borde:2008}
Ch.J.\ Bord\'e,
Eur.\ Phys.\ J.\ Spec.\ Top. {\bf 163}, 315 (2008).

\bibitem{MuellerTheseProceedings}
H.\ M\"uller, M.A.\ Hohensee, and N.\ Yu, 
these proceedings.

\bibitem{Dimopoulos:2008a}
S.\ Dimopoulos, P.W.\ Graham, J.M.\ Hogan, and M.A.\ Kasevich,
Phys.\ Rev.\ D {\bf 78}, 042003 (2008);
Phys.\ Rev.\ Lett. {\bf 98}, 111102 (2007).

\bibitem{Reasenberg:2010}
R.D.\ Reasenberg and J.D.\ Phillips,
Class.\ Quantum\ Grav. {\bf 27}, 095005 (2010).



\end{thebibliography}
\end{document}